\documentclass[reprint,amsmath,amssymb,aps]{revtex4-2}
\usepackage{CJK}
\usepackage{graphicx}
\usepackage{color}
\usepackage{amsmath}
\usepackage{amssymb}
\usepackage{amsthm}
\usepackage{extarrows}
\usepackage{tabularx}
\usepackage{dsfont}
\usepackage{booktabs}
\usepackage{float}
\usepackage[dvipdfm,colorlinks,urlcolor=blue,linkcolor=blue,anchorcolor=blue,citecolor=blue]{hyperref}
\usepackage{lineno}

\usepackage{appendix}

\begin{document}

\title{Multi-Type Instability Processes of Periodic Orbits in Nonlinear Chains}

\author{Weicheng Fu$^{1,2,3}$}
\email{fuweicheng@tsnu.edu.cn}
\author{Zhen Wang$^{4}$}
\author{Yong Zhang$^{5,3}$}
\email{yzhang75@xmu.edu.cn}
\author{Hong Zhao$^{5,3}$}

\affiliation{
$^1$ Department of Physics, Tianshui Normal University, Tianshui 741000, Gansu, China\\
$^2$ Key Laboratory of Atomic and Molecular Physics $\&$ Functional Material of Gansu Province, College of Physics and Electronic Engineering, Northwest Normal University, Lanzhou 730070, China\\
$^3$ Lanzhou Center for Theoretical Physics, Key Laboratory of Theoretical Physics of Gansu Province, and Key Laboratory of Quantum Theory and Applications of MoE, Lanzhou University, Lanzhou, Gansu 730000, China\\
$^4$ CAS Key Laboratory of Theoretical Physics and Institute of Theoretical Physics,
Chinese Academy of Sciences, Beijing 100190, China\\
$^5$ Department of Physics, Xiamen University, Xiamen 361005, Fujian, China
}

\date{\today }

\begin{abstract}
Nonlinear normal modes are periodic orbits that survive in nonlinear many-body Hamiltonian systems, and their instability is crucial for relaxation dynamics. Here, we study the instability process of the $\pi/3$-mode in the Fermi-Pasta-Ulam-Tsingou-$\alpha$ chain with fixed boundary conditions. We find that three types of bifurcations---period-doubling, tangent, and Hopf---coexist in this system, each driving instability at specific reduced wave-number $\tilde{k}$. Our analysis reveals a universal scaling law for the instability time $\mathcal{T} \propto (\lambda - \lambda_{\rm c})^{-1/2}$, independent of bifurcation types and models, where the critical perturbation strength $\lambda_{\rm c}$ scales as $\lambda_{\rm c} \propto (\tilde{k} - \tilde{k}_{\rm c})$, with $\tilde{k}_{\rm c}$ varying across bifurcations. We also observe a double instability phenomenon for certain system sizes, meaning that larger perturbations do not always lead to faster thermalization. These results provide new insights into the relaxation and thermalization dynamics in many-body systems.
\end{abstract}

\maketitle

\emph{Introduction}---Conservation laws govern the relaxation and transport dynamics of many-body Hamiltonian systems \cite{PhysRevLett.110.070604,PhysRevB.55.11029,PhysRevB.82.104306,CRPHYS2019,Complex2012Book,Lu2023}. When the number of conserved quantities equals that of degrees of freedom, the system is integrable, resulting in non-thermalization and ballistic transport behavior \cite{Lebowitz1967,lepri2003thermal,Dhar08AdvPhys,Lepri2016,ZHAO2021CPS}. Under small perturbations, most conserved quantities are destroyed, yet some persist according to the Kolmogorov-Arnold-Moser theorem \cite{Arnold1989}, as exemplified by the Fermi-Pasta-Ulam-Tsingou (FPUT) recurrence \cite{fermi1955studies,ford1992fermi,2008LNP728G}, which arises from $q$-breathers \cite{PhysRevLett.95.064102,PhysRevE.73.036618,deng2024qbreathers,Campbell2024Chaos} or $q$-tori \cite{PhysRevE.81.016210}. The breaking of invariant tori (i.e. periodic orbits)---whether, when, and how it occurs---has a critical effect on the relaxation properties of a system. For instance, if the initial state of the system lies precisely on these periodic orbits, the prerequisite for its thermalization is the breaking of these orbits, followed by resonant wave-wave interactions \cite{Onorato4208,PhysRevLett.120.144301,EPL2018Thermalization}, ultimately leading to an equilibrium state. The related subject remains an active research area, including studies on Bose gases \cite{Kinoshita2006,PhysRevLett.125.240604,Le2023}, nonlinear phononic \cite{PhysRevLett.112.075505} and photonic systems \cite{PhysRevX.4.011054,PhysRevX.8.041017}, and other quantum systems \cite{Hudomal2020,PhysRevLett.127.130601,PhysRevX.12.031026}.

Nonlinear normal modes (NNMs) act as special periodic orbits, and their instability dynamics significantly influence the system's relaxation and thermalization behavior \cite{PhysD1983, PhysRevE.54.5766, FLACH1996223, poggi1997exact, CHECHIN2002208, RINK200331, Peng_2022}. Extensive studies have shown that NNMs undergo bifurcations and become unstable when the perturbation strength $\lambda$ exceeds a critical threshold $\lambda_{\rm c}$ \cite{PhysRevE.69.046604, PhysRevE.70.016611, CHECHIN2005121, PhysRevE.73.056206, PhysRevE.74.047201, 2006IJBC, leo2007stability, chechin2012stability, PhysRevE.94.042209}. For instance, previous work demonstrated that NNMs in the FPUT-$\beta$ model destabilize via a period-doubling bifurcation, while in the Bose-Einstein condensation model, they exhibit a tangent bifurcation \cite{2006IJBC}.

In this work, we study the instability dynamics of the $\pi/3$-mode in the FPUT-$\alpha$ model with fixed boundary conditions. We show that three types of bifurcations---period-doubling, tangent, and Hopf---coexist in this system. The key to revealing this coexistence lies in analyzing the error evolution equation in the \emph{reduced wave-number} $\tilde{k}$ space, which enables a more comprehensive exploration of the parameter space and uncovers this richer, previously unobserved phenomenon. Besides, both Floquet theory and numerical simulations consistently show that the instability time $\mathcal{T}$ follows a critical exponent of $1/2$, i.e., $\mathcal{T} \propto (\lambda - \lambda_c)^{-1/2}$, independent of the bifurcation type, the NNM's wave-number, or the specific model. We also find that $\lambda_{\rm c} \propto (\tilde{k} - \tilde{k}_{\rm c})$, with $\tilde{k}_{\rm c} = 0$ for the Hopf bifurcation and $\tilde{k}_{\rm c} = \frac{2}{\pi} \arcsin(\frac{\sqrt{3}}{4})$ for the period-doubling bifurcation. Moreover, we observe a double instability phenomenon in the destabilization dynamics of the $\pi/3$-mode at certain system sizes, implying that larger perturbations do not always accelerate thermalization. This provides valuable insights into the relaxation and thermalization processes in many-body Hamiltonian systems.

\emph{The Model}---The FPUT-$\alpha$ chain consists of $N$ particles of unit mass interacting with their nearest neighbors, and its Hamiltonian reads
\begin{equation}\label{eq-Ham}
  H=\sum_{j=1}^{N}\left[\frac{p_{j}^{2}}{2}+\frac{\left(x_{j}-x_{j-1}\right)^{2}}{2}+\frac{\alpha}{3}\left(x_{j}-x_{j-1}\right)^{3}\right],
\end{equation}
where $p_j$ and $x_j$ are, respectively, the momentum and displacement from the equilibrium position of the $j$-th particle. The parameter $\alpha$ is the nonlinear coupling strength, which controls the \emph{perturbation strength} $\lambda = \alpha^2 \varepsilon$ \cite{fu2019universal}, with  $\varepsilon$ being the energy per particle.

For convenience, $x_{j}$ is represented by normal modes
\begin{equation}\label{eq-normal-modes}
x_{j}(t)=\sqrt{\frac{2}{N}} \sum_{k=1}^{N-1} Q_{k}(t) \sin \left(\frac{jk\pi}{N}\right),
\end{equation}
for fixed boundary conditions $x_0 = p_0 = x_N = p_N = 0$, where there are $N-1$ moving particles. Here, $Q_k$ represents the amplitude of the $k$-th normal mode.

By combining Eqs.~(\ref{eq-Ham}) and (\ref{eq-normal-modes}), we derive the equations of motion for the normal modes
\begin{equation}\label{eq-motion-k}
\ddot{Q}_k = -\omega_k^2 Q_k - \frac{\alpha}{\sqrt{2N}} \sum_{k_2, k_3} \omega_k \omega_{k_2} \omega_{k_3} C_{k,k_2,k_3} Q_{k_2} Q_{k_3},
\end{equation}
where $\omega_k$ is the frequency of the $k$-th mode, given by
\begin{equation}\label{eq-omega-k}
\omega_k = 2 \sin \left( \frac{\pi k}{2N} \right) = 2 \sin \left( \frac{\tilde{k} \pi}{2} \right), \quad 1 \le k \le N-1,
\end{equation}
and $\tilde{k} = k/N$ is the \emph{reduced wave-number}. Here $C_{k_1, k_2, k_3}$ represents the selection rules
\begin{equation}\label{eq-Ck123}
\begin{aligned}
C_{k_1, k_2,k_3}
=&\delta_{k_1- k_2-k_3,0}+\delta_{k_1- k_2+k_3,0}+\\
&\delta_{k_1+ k_2-k_3,0}-\delta_{k_1+ k_2+k_3,2N},
\end{aligned}
\end{equation}
where $\delta$ is the Kronecker delta function.

\emph{The solution of $\pi/3$-mode}---Suppose only the $\pi/3$-mode (i.e., $k=2N/3$, and $N$ is an integer multiple of 3) is initially excited, while all other modes remain frozen. Under this condition, Eq. (\ref{eq-motion-k}) reduces to
\begin{equation}\label{eq-motion-2n3-mode}
\ddot{Q}_{\pi/3} = -3 Q_{\pi/3} + \frac{3\sqrt{3} \alpha}{\sqrt{2N}} Q_{\pi/3}^2,
\end{equation}
whose solution can be expressed in terms of the Jacobi elliptic cosine function $\text{cn}$ as
\begin{equation}\label{eq-Q2n3}
Q_{\pi/3} = a + b\, \text{cn}^2[\Omega t, m],
\end{equation}
where $a=\frac{\theta-1+2m}{\alpha\theta}\sqrt{\frac{N}{6}}$,
$b= -\frac{3m}{\theta}\sqrt{ \frac{N}{6}}$, and $\Omega ^2=\frac{3}{4\theta}$, in which
$\theta=\sqrt{m^2 -m +1}$,
and the parameter $m=\kappa^2$ is defined by the elliptic modulus $\kappa$, which is related to the \emph{perturbation strength} by
\begin{equation}\label{eq-Energy-m}
\lambda=\frac{N}{N-1}\left[\frac{1}{6}-\frac{(m+1)\left(2m ^2-5m +2\right)}{12\left(m ^2-m +1\right)^{3/2}}\right].
\end{equation}
Note that the period $T$ of $Q_{\pi/3}$ is half the period of the cn, given by $T = {2K(m)}/{\Omega}$, where $K(m)$ is the complete elliptic integral of the first kind (see Sec. A in the Supplemental Material \cite{SM} for the solution derivation).

\emph{Linear stability analysis}---Assuming only the $\pi/3$-mode is excited, let $\mathcal{Q}_k$ represent the error in mode $Q_k$. From Eq.~(\ref{eq-motion-k}), we derive the equation of motion for $\mathcal{Q}_k$:
\begin{equation}\label{eq-D-Qt}
\ddot{\mathcal{Q}}_k = -\omega_k^2 \mathcal{Q}_k - \alpha \omega_k\sqrt{\frac{6}{N}} Q_{\frac{\pi}{3}} \sum_{k_3} \omega_{k_3} C_{k, \frac{2N}{3}, k_3} \mathcal{Q}_{k_3},
\end{equation}
where $C_{k,2N/3,k_3}$ is nonzero only when $k_3 = k - 2N/3$, $k_3 = 2N/3 - k$, $k_3 = k + 2N/3$, or $k_3 = 4N/3 - k$, which results in the following three-mode coupled equations
\begin{equation}\label{eq-motion-C-err}
\begin{cases}
\ddot{\mathcal{Q}}_{k}&=
-\omega_{k}^{2} \mathcal{Q}_{k}-
\Lambda \omega_{k}\left(
 \omega_{k_2} \mathcal{Q}_{k_2}+
 \omega_{k_3} \mathcal{Q}_{k_3}
\right),\\
\ddot{\mathcal{Q}}_{k_2}&=
-\omega_{k_2}^{2}  \mathcal{Q}_{k_2}-
\Lambda\omega_{k_2}\left(
 \omega_{k} \mathcal{Q}_{k}-
 \omega_{k_3}\mathcal{Q}_{k_3}
 \right),\\
\ddot{\mathcal{Q}}_{k_3}&=
-\omega_{k_3}^{2} \mathcal{Q}_{k_3}
-\Lambda\omega_{k_3} \left(
 \omega_{k} \mathcal{Q}_{k}-
 \omega_{k_2}\mathcal{Q}_{k_2}
 \right),
\end{cases}
\end{equation}
where $k\in(0,{N}/{3})$, $k_2={2N}/{3}-k\in({N}/{3},{2N}/{3})$,  $k_3={2N}/{3}+k\in({2N}/{3},N)$, and
\begin{equation}\label{eq-Lambda-t}
\Lambda=\alpha\sqrt{\frac{6}{N}} Q_{\frac{\pi}{3}}=
\left(1-\frac{1-2m}{\theta}\right)-\frac{3m}{\theta}~\text{cn}^2[\Omega t,m],
\end{equation}
which is a periodic function and also the driving term in Eq.~(\ref{eq-motion-C-err}) (see Sec. B in the Supplemental Material \cite{SM} for details).

Defining $\boldsymbol{W}=[\mathcal{Q}_k~~
  \dot{\mathcal{Q}}_{k}~~{\mathcal{Q}_{k_2}}
  ~~\dot{\mathcal{Q}}_{k_2}~~
  {\mathcal{Q}_{k_3}}~~\dot{\mathcal{Q}}_{k_3}]^{\rm T}$, then Eq.~(\ref{eq-motion-C-err}) can be further rewritten as
\begin{equation}\label{eq-W-matrix} \dot{\boldsymbol{W}}=\boldsymbol{A}(t)\boldsymbol{W},
\end{equation}
where the coefficient matrix
\begin{equation}\label{eq-coeff-A}
\boldsymbol{A}(t)
  =
  \begin{bmatrix}
    0 & 1 & 0 & 0 & 0 & 0 \\
    -\omega_k^2 & 0 & -\Lambda\omega_k\omega_{k_2} & 0 & -\Lambda\omega_k\omega_{k_3} & 0 \\
    0 & 0 & 0 & 1 & 0 & 0 \\
    -\Lambda\omega_k\omega_{k_2} & 0 & -\omega_{k_2}^2 & 0 & \Lambda\omega_{k_2}\omega_{k_3} & 0 \\
    0 & 0 & 0 & 0 & 0 & 1 \\
    -\Lambda\omega_k\omega_{k_3} & 0 & \Lambda\omega_{k_2}\omega_{k_3} & 0 & -\omega_{k_3}^2 & 0
  \end{bmatrix}
\end{equation}
with the period of $T$.

According to the Floquet theory \cite{Teschl2012}, the solution of Eq.~(\ref{eq-W-matrix}) has the following form
\begin{equation}\label{eq-W-muS}
 \boldsymbol{W}(t)\sim e^{\mu t}\boldsymbol{S}(t),
\end{equation}
where $\boldsymbol{S}(t)$ is periodic with period $T$, and $\mu$ is the Floquet exponent, which may be complex. The principal fundamental matrix $\boldsymbol{X}(t)$ for the system (\ref{eq-W-matrix}) is given by
\begin{equation}\label{eq-X-matrix}
  \dot{\boldsymbol{X}}(t)=\boldsymbol{A}(t)\boldsymbol{X}(t).
\end{equation}
with the initial condition $\boldsymbol{X}(0)=\boldsymbol{I}$ (i.e., unit matrix) \cite{Teschl2012}.
The eigenvalues $\rho_j$ of $\boldsymbol{X}(T)$ are known as the Floquet multipliers, which are related to the Floquet exponents $\mu_j$ by $\mu_j = {\ln(\rho_j)}/{T}$. If $ |\rho_j| \leq 1 $ (i.e., $\Re(\mu_j) \leq 0$) for all $ j \in [1, 6] $, the system is stable. Conversely, if any $ |\rho_j| > 1 $ (i.e., $\Re(\mu_j) > 0$), the system becomes unstable, as the error will grow exponentially, as shown in Eq.~(\ref{eq-W-muS}), and the instability time $\mathcal{T}$ can be estimated as
\begin{equation}\label{eq-T-mu}
  \mathcal{T} \propto \frac{1}{\Re(\mu_j)}.
\end{equation}

\begin{figure}[t]
  \centering
  \includegraphics[width=1\columnwidth]{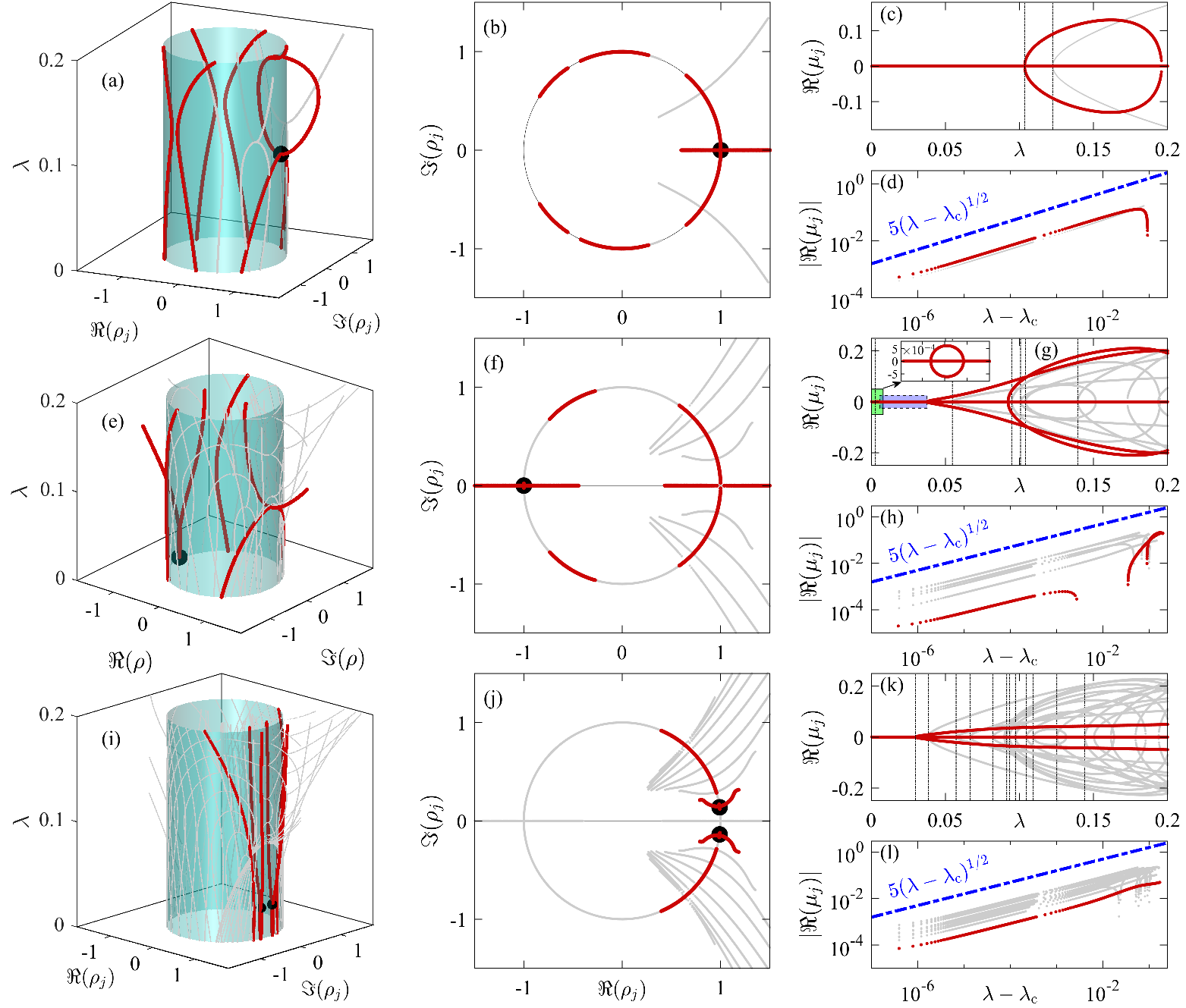}\\
  \caption{Panels (a) and (b) show how $\rho_j$s move off the unit circle as $\lambda$ increases for $N=9$. The $\rho_j$ that determines $\lambda_{\rm c}$ is plotted in red, while the others are in grey. Panel (c) shows $\Re{(\mu_j)}$ versus $\lambda$, with the dashed line marking the transition from zero to non-zero. Panel (d) shows $|\Re{(\mu_j)}|$ versus $\lambda - \lambda_{\rm c}$, with a dashed blue reference line. Panels (e)-(h) and (i)-(l) replicate (a)-(d) for $N=21$ and $39$, respectively. The inset in panel (g) zooms in on the green region where $\lambda \in [0, 0.008]$.
  }\label{fig-Circle}
\end{figure}

In practice, given $N$, we select a mode $k \in [1, N/3)$, integrate Eq.~(\ref{eq-X-matrix}) to obtain $\boldsymbol{X}(T)$, and determine the instability threshold $\lambda_{{\rm c},k}$ for the $k$-th mode by solving $D = 0$ [where $D$ represents the discriminant of the characteristic polynomial of $\boldsymbol{X}(T)$]. $D(\lambda_{{\rm c},k}) = 0$ means that a pair of $\rho_j$ collide on the unit circle and will leave (see Fig.~\ref{fig-Circle}). We define the critical threshold $\lambda_{\rm c}$ as the minimum value among $\lambda_{{\rm c},k}$, i.e., $\lambda_{\rm c} = \min\{\lambda_{{\rm c},k}\}$, above which the system becomes unstable. The corresponding mode index $k_{\rm c}$ identifies the first mode to lose stability. Next, we derive $k_{\rm c}$.

\begin{figure}[t]
  \centering
  \includegraphics[width=1\columnwidth]{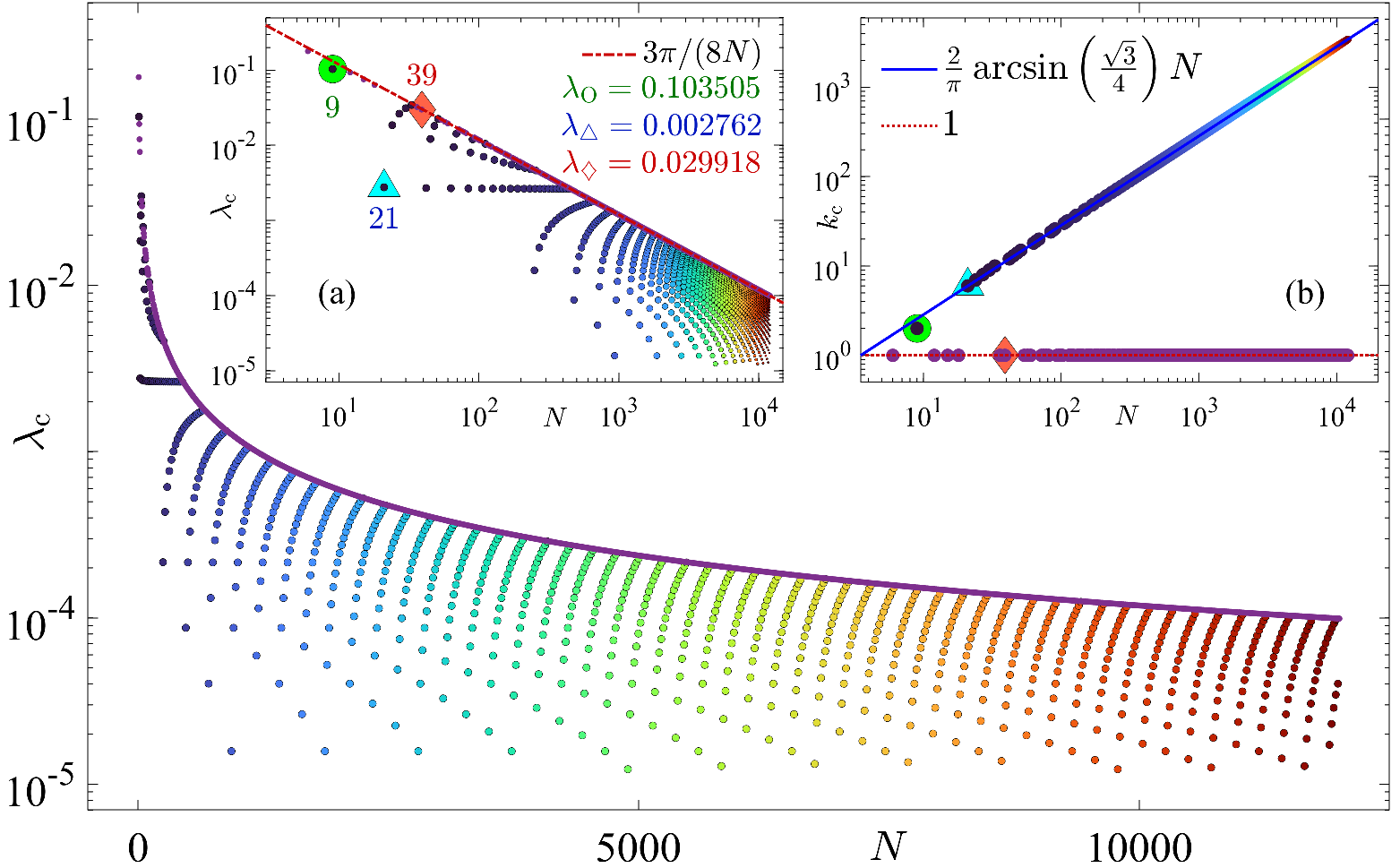}
  \caption{Dependence of $\lambda_{\rm c}$ on $N$. Inset (a): Same as the main panel but in log-log scale, with the red dashed line as a reference. The $\lambda$ with different subscripts corresponds to $\lambda_{\rm c}$ for the size indicated by the shape. Inset (b): $k_{\rm c}$ versus $N$.
  }\label{fig-lambdaC}
\end{figure}

In fact, given $\tilde{k}$, $m$ (or $\lambda$) is the sole argument of $\boldsymbol{X}(T)$. On the other hand, it is known that the NNM is \emph{linearly stable}, and, in general, $\lambda_{\rm c}$ will approach zero in the thermodynamic limit, i.e., $\lambda_{\rm c} \to 0$ (viz. $m \to 0$) as $N \to \infty$. Therefore, we reverse the approach: let $m \to 0$, then ${\rm cn}(\Omega t, m) \to \cos(\Omega t)$, and now Eq.~(\ref{eq-X-matrix}) depends only on $N$ (or more precisely, on $\tilde{k}$), making the equation solvable. Solving $D(\tilde{k}_{\rm c}) = 0$ will determine which mode becomes unstable first. If there is no solution for $D(\tilde{k}_{\rm c}) = 0$, this implies that the assumption is invalid, meaning that in the thermodynamic limit, $\lambda_{\rm c}$ is \emph{finite}, \emph{not zero}. Here, for $m\to0$, we obtain
\begin{equation}\label{eq-eta2k}
\tilde{k}_{\rm c}=\frac{k_{\rm c}}{N}=
\begin{cases}
  0,\quad \text{tangent or Hopf bifurcation};\\
  \frac{2}{\pi}\arcsin\left(\frac{\sqrt{3}}{4}\right),~\text{period-doubling},
\end{cases}
\end{equation}
which implies that the mode $k_{\rm c} = 1$ or $k_{\rm c} = \lfloor N \tilde{k}_{\rm c} \rfloor$ will lose stability first, where $\lfloor \cdot \rfloor$ denotes the integer part (see Sec. C in the Supplemental Material \cite{SM} for details, especially discussions for bifurcation's type).

Although an \emph{approximate solution} to Eq.~(\ref{eq-X-matrix}) can be obtained using perturbation theory \cite{leo2007stability}, the resulting expression is too lengthy to provide clear physical intuition. To gain better insight, we instead solve the equation numerically using the embedded Runge-Kutta-Nystrom algorithm of order 12(10) \cite{dormand1987high} and present the results graphically.

\emph{Numerical solutions}---We first calculate $\lambda_{\rm c}$ and $k_{\rm c}$ for different system sizes $N$, as shown in Fig.~\ref{fig-lambdaC}. The scatter points represent the values of $\lambda_{\rm c}$ corresponding to $k_{\rm c}$, revealing a rich structure. Inset (a) displays the same results as the main panel, but in a log-log scale. Inset (b) shows the dependence of $k_{\rm c}$ on $N$, where two distinct behaviors are observed: one with $k_{\rm c} = 1$ and the other with $k_{\rm c} = \lfloor \frac{2N}{\pi} \arcsin(\frac{\sqrt{3}}{4})\rfloor$, confirming the result in Eq.~(\ref{eq-eta2k}). Additionally, we find that $\lambda_{\rm c} \propto N^{-1}$ when $k_{\rm c} = 1$, as indicated by the red dashed line in inset (a), which suggests $\lambda_{\rm c} \propto \tilde{k}$ in terms of the reduced wave number.

\begin{figure}[t]
  \centering
  \includegraphics[width=1\columnwidth]{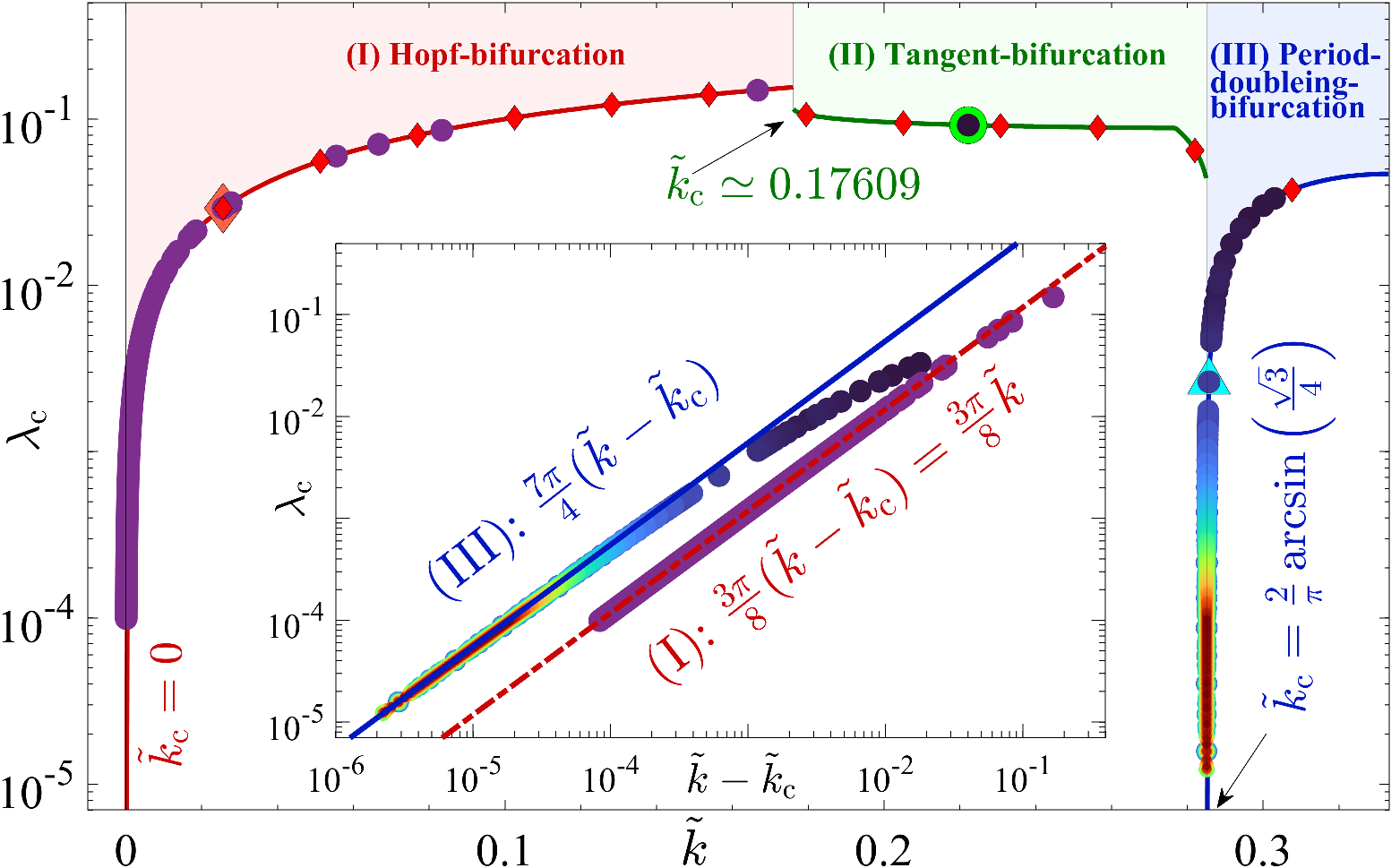}
  \caption{Dependence of $\lambda_{\rm c}$ on $\tilde{k}$. The red diamonds represent $\lambda_{{\rm c},k}$ for $N=39$, with $k \in [1,12]$. Inset: Same as the main panel [regions (I) and (III)], but with the horizontal axis replaced by $(\tilde{k} - \tilde{k}_{\rm c})$. Straight lines are shown for reference.
  }\label{fig-lambdaC_eta}
\end{figure}

Figure~\ref{fig-lambdaC_eta} shows $\lambda_{\rm c}$ as a function of $\tilde{k}$. The seemingly irregular data from Fig.~\ref{fig-lambdaC} are now clearly situated on three distinct curves. The solid curves are obtained by solving Eq.~(\ref{eq-X-matrix}) for $\tilde{k}$, treating it as a continuous variable in the range $[0, 1/3)$, without accounting for the system's discreteness and its constraints on $\tilde{k}$. The discrete results from Fig.~\ref{fig-lambdaC} are rescaled by $(1 - 1/N)$ for comparison with the solid curve, as shown by the circles. Clearly, the solid curve displays discontinuities, dividing $[0, 1/3)$ into three regions, which from left to right correspond to the Hope bifurcation (I), tangent bifurcation (II), and period-doubling bifurcation (III), as evidenced in Fig.~\ref{fig-Circle}. Region (II) corresponds to a finite $\lambda_{\rm c}$ in the range $(0.044, 0.116)$, and for $N \in [6, 12000]$, only the $\lambda_{\rm c}$ for $N=9$ falls within this range. Moreover, we see that $\tilde{k}_{\rm c} = 0$ for region (I) and $\tilde{k}_{\rm c} = \frac{2}{\pi} \arcsin(\frac{\sqrt{3}}{4})$ for region (III), which are in good agreement with Eq.~(\ref{eq-eta2k}). To observe the scaling behavior of $\lambda_{\rm c}$ as $\tilde{k}$ approaches $\tilde{k}_{\rm c}$ in these two regions, we plot $\lambda_{\rm c}$ as a function of $(\tilde{k} - \tilde{k}_{\rm c})$ in the inset. The best linear fit suggests that $\lambda_{\rm c}$ approximately follows the relation
\begin{equation}\label{eq-lambdaC}
\lambda_{\rm c} \simeq
\begin{cases}
  \frac{3\pi}{8}(\tilde{k} - \tilde{k}_{\rm c}), & \text{(I)}:~\tilde{k}_{\rm c}=0;\\
  \frac{7\pi}{4}(\tilde{k} - \tilde{k}_{\rm c}), & \text{(III)}:~\tilde{k}_{\rm c}=\frac{2}{\pi}\arcsin\left(\frac{\sqrt{3}}{4}\right),
\end{cases}
\end{equation}
for $\tilde{k}$ approaching $\tilde{k}_{\rm c}$, as shown by the lines in the inset.

For a given $N$, $\lambda_{{\rm c},k}$ follows the same pattern as presented above, with $k \in [1, N/3)$. As shown by the red diamonds for $N=39$ in Fig.~\ref{fig-lambdaC_eta}. Considering Eq.~({\ref{eq-omega-k}}), it is concluded that different frequencies in the system correspond to different bifurcation types: $\omega_k \in (0, 0.5462)$ for region (I), $\omega_k \in [0.5462, 0.866)$ for region (II), and $\omega_k \in [0.866, 1.7321)$ for region (III). This highlights the strong predictive power of Fig.~\ref{fig-lambdaC_eta}: for any $N$, $\lambda_{{\rm c},k}$ for $\tilde{k}$ can be directly read from the plot. The minimum value corresponds to the system's $\lambda_{\rm c}$ and also identifies the type of bifurcation responsible for the instability.

\begin{figure}[t]
  \centering
  \includegraphics[width=1\columnwidth]{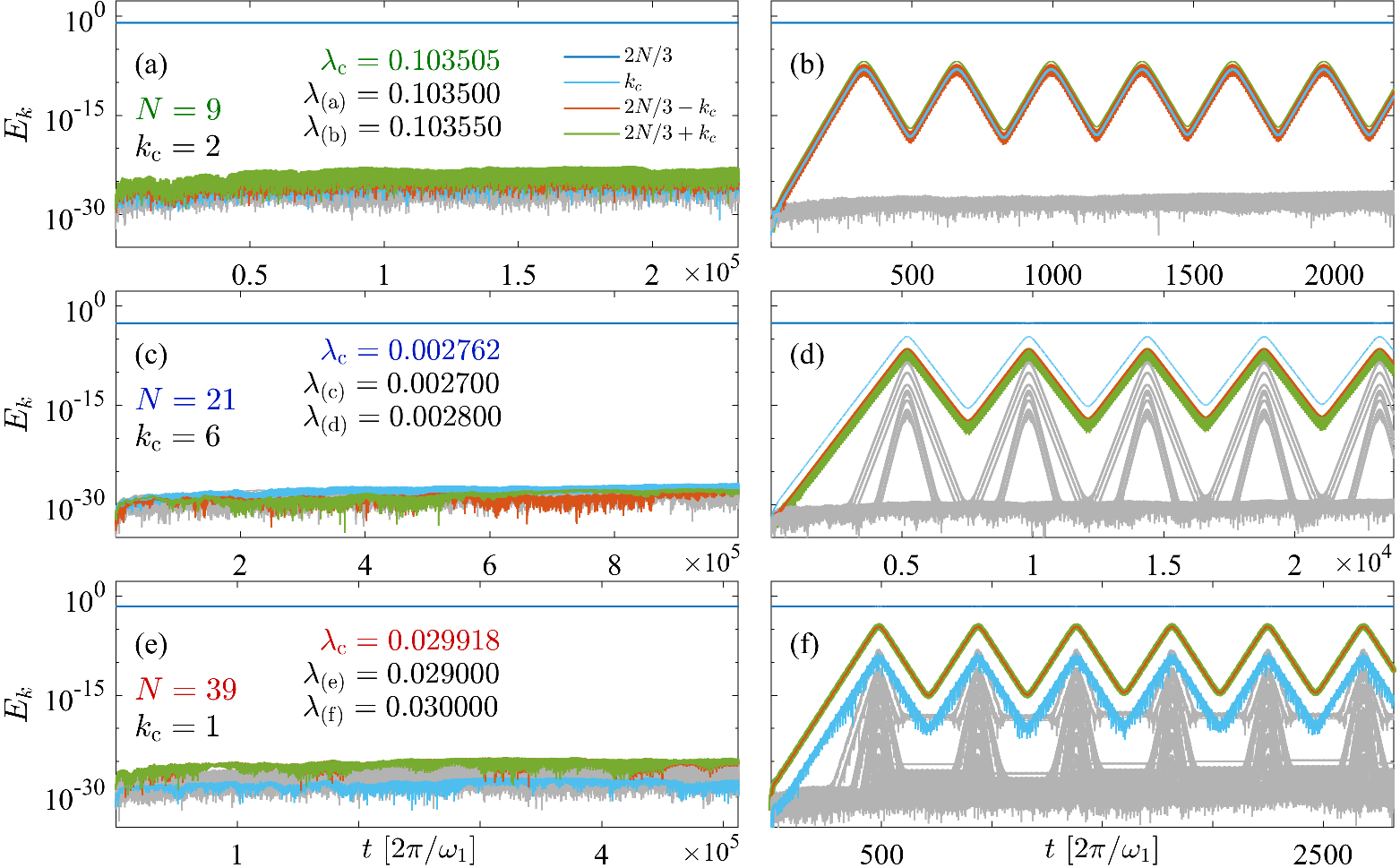}
  \caption{Panels (a) and (b) show the evolution of $E_k$ for $N=9$, with $\lambda_{\rm (a)}$ and $\lambda_{\rm (b)}$ corresponding to the results in each panel, respectively. Panels (c) and (d) show results for $N=21$, while panels (e) and (f) are for $N=39$.
  }\label{fig-Qt}
\end{figure}

\begin{figure}[t]
  \centering
  \includegraphics[width=1\columnwidth]{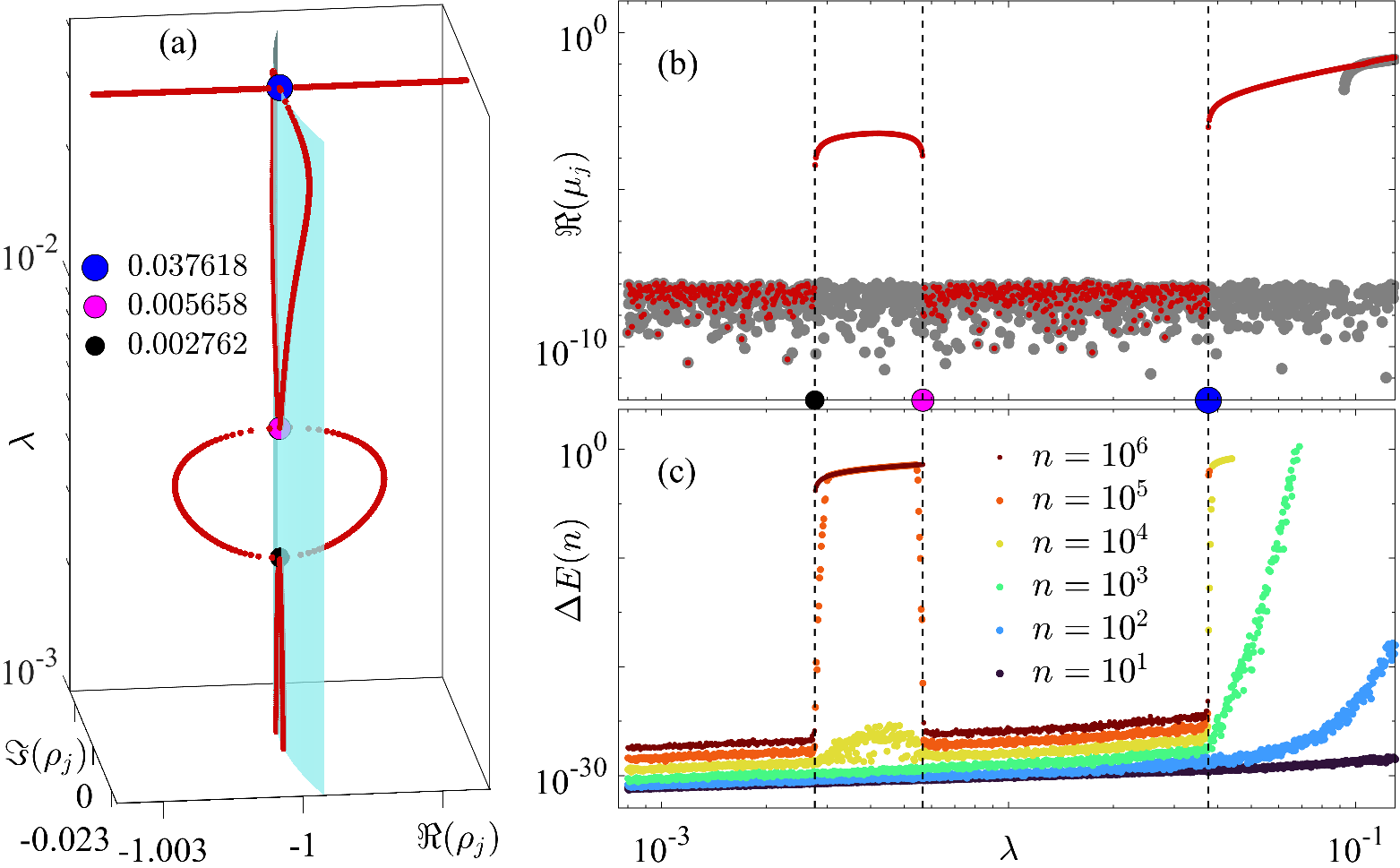}\\
  \caption{Panel (a) shows how $\rho_j$ shifts off the unit circle as $\lambda$ increases for $N=21$. Panel (b) shows $\Re{(\mu_j)}$ versus $\lambda$. Panel (c) shows $\Delta E(n)$ versus $\lambda$ at different times $n$.
  }\label{fig-Errs}
\end{figure}

We now validate the accuracy of Floquet theory for threshold's prediction through molecular dynamics simulations (MDS). Initially, the total energy is assigned to the $2N/3$ mode, and the particles' positions are initialized using Eq.~(\ref{eq-normal-modes}). The Hamiltonian canonical equations, derived from Eq.~(\ref{eq-Ham}), are numerically evolved with the same algorithm. The energy of each mode, $E_k(t)$, is monitored, as shown in Fig.~\ref{fig-Qt}, where $E_k$ is defined as
\begin{equation}\label{eq-EnKt}
E_{k}=
\begin{cases}
\frac{1}{2}\left(P_{k}^{2}+\omega_{k}^{2} Q_{k}^{2}\right),&k\neq 2N/3;\\
\frac{1}{2} \left( P_k^2 + 3 Q_{k}^2 \right) - \alpha \sqrt{\frac{3}{2N}} Q_{k}^3,&k= 2N/3.
\end{cases}
\end{equation}
Here, $P_k=\partial H/\partial\dot{Q}_k$ is the canonical conjugate momentum. For $\lambda < \lambda_{\rm c}$ (left panels), $E_{2N/3}(t)$ remains nearly constant, while the energy in other modes stays close to zero (within numerical precision), indicating stability.  In contrast, for $\lambda > \lambda_{\rm c}$ (right panels), energy in the error-coupled modes grows exponentially, and the system exhibits quasi-periodic oscillations. Yet, the system finally becomes unstable, as shown in Fig.~\ref{fig-Errs}(c). These results agree well with our theoretical predictions. We now proceed to examine the types of bifurcations.

Figures~\ref{fig-Circle}(a) and \ref{fig-Circle}(b) show the dependence of $\rho_j$ on $\lambda$ for $N = 9$. A pair of $\rho_j$ first depart from the unit circle at $+1$, indicating a tangent bifurcation that destabilizes the system. The gray line corresponds to $k = 1$, where a pair of complex conjugate eigenvalues leave the unit circle, signaling a Hopf bifurcation. Figure~\ref{fig-Circle}(c) shows $\Re(\mu_j)$ as a function of $\lambda$, clearly illustrating that $\lambda_{\rm c} = \lambda_{{\rm c},2} < \lambda_{{\rm c},1}$. Since $\rho_j$ are the roots of the characteristic polynomial of $\boldsymbol{X}(T)$, and $\mu_j = \ln(\rho_j)/T$, we have $\Re(\mu_j(\lambda_{\rm c})) = 0$. Thus, the lowest-order expansion of $\Re(\mu_j)$ near $\lambda_{\rm c}$ is expected to scale as $(\lambda - \lambda_{\rm c})^{1/2}$, consistent with the numerical results in Fig.~\ref{fig-Circle}(d). As a result, the instability time $\mathcal{T}$ near $\lambda_{\rm c}$ scales as $(\lambda - \lambda_{\rm c})^{-1/2}$.

Figures~\ref{fig-Circle}(e)-\ref{fig-Circle}(h) show the results for $N = 21$, where a pair of $\rho_j$ first depart from the unit circle at $-1$, indicating a period-doubling bifurcation. Figures~\ref{fig-Circle}(i)-\ref{fig-Circle}(l) show the results for $N = 39$, where a pair of $\rho_j$ leave the unit circle as complex conjugate eigenvalues, signaling a Hopf bifurcation. Additionally, the results in Figs.~\ref{fig-Circle}(g), \ref{fig-Circle}(h), and \ref{fig-Circle}(l) are qualitatively identical, confirming that $\mathcal{T} \propto (\lambda - \lambda_{\rm c})^{-1/2}$, regardless of the bifurcation type. This scaling law is further supported by the data, including the gray dots in panels (g), (h), and (l). The observed scaling matches that of the $\pi/4$-mode in the FPUT-$\alpha$ and FPUT-$\beta$ models \cite{Peng_2022}, indicating that it holds universally across different wave numbers and models. Moreover, except for $N = 9$, all three types of bifurcations are observed for other values of $N$.

Moreover, Figs.~\ref{fig-Circle}(e) and \ref{fig-Circle}(g) reveal that $\lambda_{\rm c}$ is very small. The inset in Fig.~\ref{fig-Circle}(g) clearly shows that $\Re(\mu_j) > 0$. Notably, the non-zero $\Re(\mu_j)$ curve forms an approximate circle, indicating a stability transition: as $\lambda$ increases, the system first loses stability and then enters a stable region (marked by the blue rectangle). Subsequently, further increases in $\lambda$ induce a secondary instability, constituting a \emph{double instability phenomenon}. Both instabilities originate from period-doubling bifurcations, as evidenced in Fig.~\ref{fig-Errs}(a), where $\rho_j$ pairs leave the unit circle at $-1$ twice.

Figure~\ref{fig-Errs}(b) shows $\Re(\mu_j)$ as a function of $\lambda$ in a log-log scale. Figure~\ref{fig-Errs}(c) displays the results of MDS , where $\Delta E(n) = \max\left\{\sum_{k=1, k \neq 2N/3}^{N-1} E_k(t) \mid t = 1, 2, \dots, n\right\}$. In the region where $\Re(\mu_j) > 0$ (specifically, $10^{-7}$), $\Delta E$ grows rapidly over time, suggesting that the $\pi/3$-mode will eventually lose stability. The MDS results align perfectly with the predictions from Floquet theory, as shown by the vertical dashed lines in panels (b) and (c).

\emph{Summary and discussion}---We have studied the instability dynamics of the $\pi/3$-mode in FPUT-$\alpha$ chains with fixed boundary conditions and identified four key results: (i) Three types of bifurcations---period-doubling, tangent, and Hopf---coexist, each triggering instability. (ii) The instability time $\mathcal{T}$ follows a universal scaling law, $\mathcal{T} \propto (\lambda - \lambda_{\rm c})^{-1/2}$, independent of bifurcation type, NNM's wave-number and model. (iii) The critical threshold $\lambda_{\rm c}$ scales as $\lambda_{\rm c} \propto (\tilde{k} - \tilde{k}_{\rm c})$, where $\tilde{k}_{\rm c} = 0$ for the Hopf bifurcation and $\tilde{k}_{\rm c} = \frac{2}{\pi} \arcsin(\frac{\sqrt{3}}{4})$ for the period-doubling regime. (iv) For certain system sizes, the instability dynamics exhibit a double instability phenomenon, meaning that larger perturbations do not always result in faster thermalization.

It is worth noting that the total number of NNMs is limited, so as $N$ increases, the probability of selecting NNMs as initial excitations becomes negligible. Besides, the instability threshold also tends to zero. Consequently, the thermalization dynamics of a general Hamiltonian system in this scenario will predominantly be governed by resonant wave-wave interactions \cite{Onorato4208,PhysRevLett.120.144301,EPL2018Thermalization}, leading to a universal thermalization behavior where the thermalization time scales inversely with the square of the perturbation strength (i.e., a simple power-law relationship) \cite{fu2019universal, Fu2019PRER, Pistone2018, PhysRevE.100.052102, PhysRevLett.124.186401, PhysRevE.104.L032104, Feng_2022, ONORATO20231, PhysRevLett.132.217102, Wang_2024}. However, for finite, particularly small $N$, the probability of selecting NNMs as initial excitations increases, and the instability threshold rises accordingly. In such cases, if the initial excitation corresponds to NNMs, the system exhibits a thermalization threshold: below this threshold, thermalization does not occur, while above it, both instability and multi-wave resonance mechanisms determine the dynamics. Hence, the thermalization time as a function of perturbation strength will deviate from the simple power-law behavior.

Our research uncovers a diverse range of destabilization dynamics, which not only deepen the understanding of how the system relaxes to equilibrium but also offer valuable insights into related areas of quantum systems \cite{PhysRevX.13.031013}, such as the study of (Floquet) time crystals \cite{PhysRevLett.109.160401,PhysRevLett.116.250401,PhysRevA.105.043302,PhysRevLett.129.063902,PhysRevLett.129.133001,RevModPhys.95.031001}.

\begin{acknowledgments}
This work was supported by the National Science Foundation of China (Grants No.~12465010, No.~12247106, No.~12005156, No.~11975190, and No.~12247101). W. Fu also acknowledges support from the Youth Talent (Team) Project of Gansu Province, the Long-yuan Youth Talents Project of Gansu Province, the Fei-tian Scholars Project of Gansu Province, the Leading Talent Project of Tianshui City, the Innovation Fund from the Department of Education of Gansu Province (Grant No.~2023A-106), and the Open Project Program of Key Laboratory of Atomic and Molecular Physics $\&$ Functional Material of Gansu Province (6016-202404).
\end{acknowledgments}

\bibliography{nnmRefs}

\end{document}